\newcommand{\abs}[1]{\left\| #1 \right\|}

\newcommand{\p}[1]{\left(#1\right)}
\newcommand{\br}[1]{\left[#1\right]}

\newcommand{\ba}[1]{\begin{eqnarray} #1\end{eqnarray}}

\newcommand{\be}[1]{\begin{equation} #1\end{equation}}

\newcommand{\beq}{\begin{equation}}
\newcommand{\eeq}{\end{equation}}
\def\comment#1{}

\def\id{\mathds{1}}

\documentclass[preprint,12pt]{elsarticle}

\usepackage{graphicx}
\usepackage{slashed,bm,dsfont}
\newcommand{\dslash}{\slashed{D} }
\usepackage{booktabs}
\usepackage[colorlinks=true,backref=false,linktocpage=true,
citecolor=blue,urlcolor=blue,linkcolor=blue,pdfpagemode=UseOutlines]{hyperref}

\hypersetup{%
  bookmarksnumbered=true,
  pdftitle = {},
  pdfsubject = {},
  pdfauthor = {A.~Alexandru},
  pdfkeywords = {}
}

\usepackage{amssymb}
\usepackage{amsfonts}
\usepackage{amsmath}
\usepackage{amssymb}
\usepackage{amsthm}
\usepackage{multirow}

\usepackage{algpseudocode}
\usepackage{algorithm}

\journal{Journal of Computational Physics}

\begin{document}

\begin{frontmatter}



\title{Multi-mass solvers for lattice QCD on GPUs}


\author{A. Alexandru} 
\ead{aalexan@gwu.edu}
\author{C. Pelissier} 
\ead{craigp@gwmail.gwu.edu} 
\author{B. Gamari}
\ead{bgamari@physics.umass.edu} 
\author{F. Lee} 
\ead{fxlee@gwu.edu}
\address{Department of Physics, The George Washington University, Washington, DC 20052}

\begin{abstract}
Graphical Processing Units (GPUs) are more and more frequently used for lattice QCD calculations. 
Lattice studies often require computing the quark propagators for several masses.
These systems can be solved using multi-shift inverters but these algorithms are memory
intensive which limits the size of the problem that can be solved using GPUs. In this paper, 
we show how to efficiently use a memory-lean single-mass inverter to solve multi-mass problems. 
We focus on the BiCGstab algorithm for Wilson fermions and show that the single-mass inverter not only
requires less memory but also outperforms the multi-shift variant by a factor of two.
\end{abstract}


\end{frontmatter}


\section{Introduction}
\label{intro}

\begin{figure}[t] 
\centering 
\includegraphics[width=4.5in]{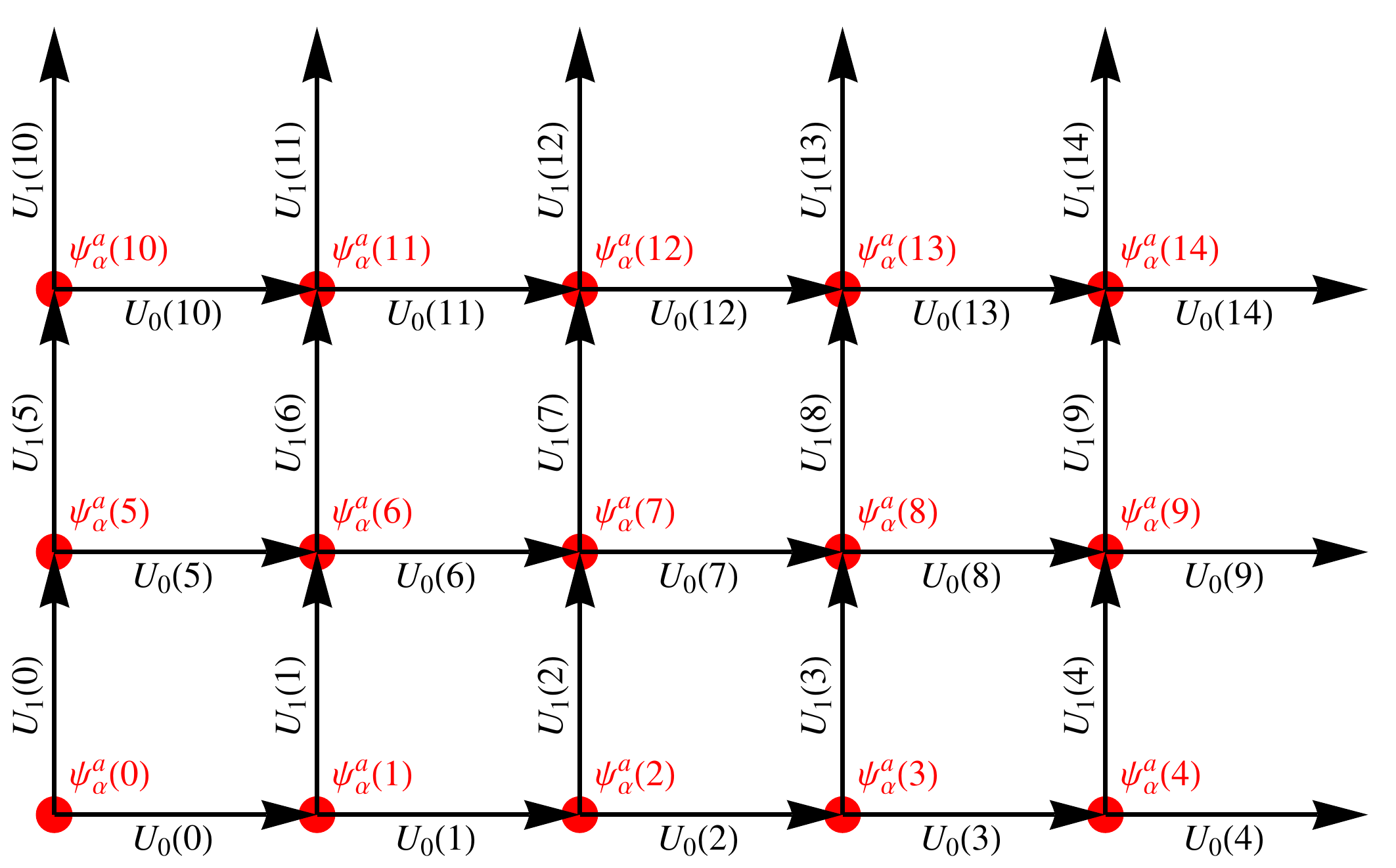} 
\caption{Schematic representation
of the lattice discretization: the quark fields $\psi^a_\alpha$ are
associated with the sites of the lattice and the gauge variables
$U_\mu$ are defined on the links connecting the sites.} 
\label{fig:lattice}
\end{figure}

While most of the visible matter in the universe is made up of
hadrons, particles that experience the strong nuclear force, their
structure is still not completely understood. The current understanding
is that hadrons are made of quarks that interact via gluons.  Quantum
chromodynamics (QCD) is the theory that describes their interactions
\cite{Politzer:1973fx,Gross:1973id}.  The structure of hadrons is determined
by a quantum superposition of many quark-gluon configurations: there
is no obvious hierarchy among the configurations and the standard
perturbative approach is not useful. However, we can use numerical
methods to compute hadron properties. These methods are based on a
non-perturbative formulation of QCD, lattice QCD
\cite{Wilson:1974sk,Wilson:1975id}: the space-time is approximated
by a four dimensional grid, the quarks are viewed as particles
hopping between the sites of this grid and the gluons are represented
by parallel transporters that change the internal state of the
quarks as they hop along the given link (see Fig. \ref{fig:lattice}
for a schematic representation). The quantum fluctuations are taken
into account by path integral methods in the Euclidean framework.

Formally, the problem is equivalent to a statistical mechanics problem
where the observables can be expressed as correlation functions; 
we use Monte Carlo techniques to estimate these correlations.
For definitiveness sake, we write down the equivalent partition function. 
After rotating to Euclidean times and discretizing the action we 
get \cite{Montvay:1994cy}:
\beq
Z=\int {\cal D} U  {\cal D} \bar{\psi} {\cal D}\psi e^{-S_g(U) - 
\bar{\psi} M(m;U) \psi}.
\eeq
The measure ${\cal D} U \equiv \prod_{n,\mu} dU_\mu(n)$ and 
${\cal D}\bar{\psi}{\cal D}\psi \equiv \prod_{n,a,\alpha}
d\bar{\psi}^a_{\alpha}(n)d\psi^a_{\alpha}(n)$ represents an integral
over all degrees of freedom: $U_\mu(n)$ is the gauge link at site
$n$ and direction $\mu$ which represents the gluon field and
$\psi^a_\alpha(n)$ is the quark field at site $n$ with color index
$a$ and spinorial index $\alpha$. The purely gluonic
part of the action, $S_g(U)$, includes the gluons' kinetic term and the gluon-gluon interaction.
The quark contribution is given by $\bar{\psi}M(m;U)\psi$ where $m$ is
the mass of the quark. Since the quark action is quadratic in the 
quark fields we can integrate them out to get
\beq 
Z=\int {\cal D} U e^{-S_g(U)} \det M(m;U).  \label{eq:2}
\eeq 
The path integral is now expressed completely in terms of the
gauge fields $U$, and the quark contribution is given by the determinant
of the quark matrix $M(m;U)$; this matrix is the discretized version
of a covariant derivative, and we will discuss its structure later
in more detail.

Most lattice QCD calculations are usually separated into two steps:
gluonic links generation and correlator measurement.  In the first
step a set of gauge configurations are generated 
with a distribution given by the Euclidean action in Eq.~\ref{eq:2};
each configuration corresponds to a snapshot of the gluonic fields
-- the quark fields are present in this stage only implicitly by
their contribution to the probability measure, $\det M(m;U)$. In the
second stage the hadron correlators are computed for each configuration,
and then the ensemble average is calculated; the hadron correlators
measure the hadron's ability to propagate through space-time. Their
behavior allows us to determine the internal structure of hadrons and
their masses. In both stages, the most time
consuming part of the calculation relates to manipulating the quark
matrix~$M(m;U)$.
The typical calculation
involves computing $M(m;U)^{-1} \bm{b}$ thousands of times for
different vectors $\bm{b}$ and gauge links $U$. The
calculation is done iteratively and most of the computation time
is spent computing the product between $M(m;U)$ and intermediate
vectors. This is referred to as the fermion-matrix multiplication routine.

Lattice calculations are numerically very expensive: the
fermionic matrix is a square matrix with a few million rows.
Fortunately, the matrix is very sparse and only a few non-zero
elements need to be stored for every row. 
Since the matrix is sparse and local, the fermion-matrix multiplication
routine can be efficiently parallelized; optimized implementations
can run double precision calculations at a rate of 1-2 GFlops per
processor core on modern CPUs \cite{Wettig:2005zc}. The numerical
requirements for lattice QCD calculations demand the use of large
clusters: typically a few hundred to a few thousand processor cores are
needed.
A new direction in high performance computing has recently become
available with the development of general purpose graphic processor
units (GPUs). These devices have very good floating point performance,
and more importantly for lattice QCD applications, very good
memory bandwidth. While these devices are difficult to program,
their raw performance is compelling. 
The advantage of using GPUs for Lattice QCD calculations was
demonstrated very early~\cite{Egri:2006zm}, and today we have
efficient implementations of the fermion-matrix multiplication
routine and of the iterative methods that allow us to compute
$M(m;U)^{-1} \bm{b}$~\cite{Clark:2009wm}.

Lattice QCD simulations often requires the computation of {\em quark
propagators} 
$M(m_i; U)^{-1} \bm{b}$ for several quark masses $m_i$.
This is needed, for example,  to study the dependence of physical 
observables on the quark mass which is required in order to
extrapolate to the physical limit (direct simulations are still
too expensive). This is also required when we 
evaluate functions of the quark matrix using rational 
approximations~\cite{Clark:2006fx,Alexandru:2007bb}.
The fermion matrices for different quark masses differ only
by a multiple of the identity matrix. They are {\em shifted} versions
of the same matrix. We can then use multi-shift 
inverters~\cite{Jegerlehner:1996pm}, iterative methods
that allow us to compute all quark propagators at once. The
fermion-matrix multiplications are only performed once (for the
smallest mass) and these results are used to compute all other
propagators. The disadvantages of these methods are that they need more
memory, require extra linear algebra operations and
common acceleration techniques cannot be used. 

In order to efficiently use GPUs we need to store most of the data
required for computation in the GPU's own memory. This is typically a few
gigabytes which is two orders of magnitude less than the memory
usually available when running lattice QCD simulations on CPU
clusters.  Thus, for GPUs we are forced to use single-mass inverters,
even when using parallel versions of our codes~\cite{Babich:2010mu,gwu:2011}.
For optimal performance we need to use as few GPUs as possible
when parallelizing lattice QCD codes.

In this paper we show how to efficiently solve multi-mass problems for lattice QCD 
using optimized  single-mass inverters. We will focus our attention on the Wilson
version of the fermionic matrix \cite{Wilson:1975id}. The plan of the paper
is the following: in Section~\ref{sec:2} we describe the numerical properties of the 
Wilson fermion matrix, in Section~\ref{sec:3} we describe the relevant GPU 
architectural details and in Section~\ref{sec:4} and \ref{sec:5} we describe our GPU 
implementation of the Wilson-matrix multiplication, the optimizations 
we used and the challenges we have encountered. In Section \ref{sec:6} 
we discuss the single-mass and multi-shift versions of the 
BiCGstab algorithm \cite{vorst:631} and the optimization techniques we used.  
In Section~\ref{sec:res} we compare the GPU performance of these two solvers.
We find that the solver
based on the single-mass inverter, not only uses less memory, but also 
outperforms the multi-shift inverter by more than a factor of two.

\section{Wilson fermions}
\label{sec:2}

Lattice QCD simulations spend most of the time in the Wilson $\dslash$
routine, a routine that multiplies a lattice vector with the Wilson
version of the fermionic matrix \cite{Wilson:1975id}. In this section
we will describe the Wilson $\dslash$ operator, discuss its implementation
and present a standard optimization technique.

The Wilson fermionic matrix is a discretized version of the continuum
fermionic operator $m+\gamma_\mu D_\mu$, where $D_\mu=\partial_\mu
+ A_\mu$ is the covariant derivative associated with the gauge field
$A_\mu$. For the definition of these operators and other details
the reader should consult a lattice field theory textbook (for example
\cite{Montvay:1994cy}) -- here we will focus on the numerical aspects
of the discretization.

The operator acts on vector functions defined on a four dimensional
space. We will only consider the values of these functions
at the grid points and we define $\psi_{n} = \psi(n a)$ where $a$
is the lattice spacing and $n$ is a 4-vector of integers used to index
the grid points. The discretized operator, $M(m; U)$, is then
\begin{eqnarray}
\label{eq:3}
\begin{split}
M(m; U)\bm{\psi}&=\left[(m a+4)\id-\frac{1}{2} \dslash(U) \right]\bm{\psi} \\
&=(m a +4)\bm{\psi} - \frac{1}{2} \sum_{\mu=\pm1,...,\pm4} T_{\mu}(U)\bm{\psi} 
\end{split}
\end{eqnarray}
where $m$ is the quark mass and $\bm{\psi}$ is the vector associated with 
$\psi_n$ (we will use bold letters to indicate lattice vectors).  
The fields that describe the quarks
are 12 dimensional: they are 4-dimensional spinors with a three
dimensional color degree of freedom. $\psi_n$ is represented by 12
complex numbers at each site; the 12 complex numbers can be organized
as a $3\times 4$ matrix: $\psi_n = ( \psi_n^{c,s} )$, with the row
index running over colors and the spinorial index over columns.  To
compute the value of the field $M(m;U)\bm{\psi}$ at one particular point
we need the value of $\bm{\psi}$ at the same point and also its value
at the neighboring points; in a four dimensional space we have eight
neighbors, two in each direction. The values of $\bm{\psi}$ at the
neighboring points are not just copied but rather {\em parallel
transported}, an operation which we denoted by $T_\mu(U)$: 
\begin{eqnarray}
\label{eq:4}
\mu>0:\quad(T_\mu \psi)_{n} &=& U_\mu(n) \psi_{n+\hat\mu} (1-\gamma_\mu) \\ \nonumber
\mu<0:\quad
(T_\mu \psi)_{n} &=& U_\mu(n-\hat\mu)^\dagger \psi_{n-\hat\mu} (1+\gamma_\mu),
\end{eqnarray}
where $\mu$ represents a direction and $n\pm\hat{\mu}$ represents
the forward/backward neighbor of $n$ in the direction $\mu$.  The $U$
matrices are $3\times 3$ special unitary matrices. For every site
$n$ on the lattice we have 4 such {\em color} matrices, one for
each direction. Multiplying the $\psi_n$ matrix from the right we
have $4\times 4$ {\em spinor} matrices which depend on the direction
but not on the lattice position. For $\gamma$-matrices we use:
\begin{equation}
\gamma_i = 
\begin{pmatrix} 
0 & i \sigma_i \cr
-i \sigma_i & 0 \cr
\end{pmatrix} \quad \mbox{for i}\in \{1,2,3\},  \quad 
\gamma_4 = 
\begin{pmatrix} 
0 & -1 \cr
-1 & 0 \cr
\end{pmatrix}, 
\end{equation}
where $\sigma_i$ are the $2\times2$ Pauli matrices.

We stress again that while we have only eight different spinor
matrices ($1\pm\gamma_\mu$), the color matrices differ from site
to site.  In fact, any optimized implementation has to take into
account the fact that the color matrices need to be transported from
memory to the processing unit; their memory layout affects the
performance of the code.

\begin{table}[t]
 \centering
 \begin{tabular}{@{} lll @{}} 
 \toprule
 Direction & vectors \\
   \midrule
   $T_1^+$      & $v_1^\dagger = \begin{pmatrix} i & 0 & 0 & 1\end{pmatrix}$  & $v_2^\dagger = \begin{pmatrix} 0& i& 1& 0\end{pmatrix}$  \\
   \midrule 
   $T_1^-$      & $v_1^\dagger = \begin{pmatrix} -i & 0 & 0 & 1\end{pmatrix}$ & $v_2^\dagger = \begin{pmatrix} 0& -i& 1& 0\end{pmatrix}$  \\
   \midrule
   $T_2^+$      & $v_1^\dagger = \begin{pmatrix} -1 & 0 & 0 & 1\end{pmatrix}$ & $v_2^\dagger = \begin{pmatrix} 0& 1& 1& 0\end{pmatrix}$  \\
   \midrule 
   $T_2^-$      & $v_1^\dagger = \begin{pmatrix} 1 & 0 & 0 & 1\end{pmatrix}$ & $v_2^\dagger = \begin{pmatrix} 0& -1& 1& 0\end{pmatrix}$  \\
   \midrule
   $T_3^+$      & $v_1^\dagger = \begin{pmatrix} 0 & -i & 0 & 1\end{pmatrix}$  & $v_2^\dagger = \begin{pmatrix} i& 0& 1& 0\end{pmatrix}$  \\
   \midrule 
   $T_3^-$      & $v_1^\dagger = \begin{pmatrix} 0 & i & 0 & 1\end{pmatrix}$ & $v_2^\dagger = \begin{pmatrix} -i& 0& 1& 0\end{pmatrix}$  \\
   \midrule
   $T_4^+$      & $v_1^\dagger = \begin{pmatrix} 0 & 1 & 0 & 1\end{pmatrix}$ & $v_2^\dagger = \begin{pmatrix} 1& 0& 1& 0\end{pmatrix}$  \\
   \midrule 
   $T_4^-$      & $v_1^\dagger = \begin{pmatrix} 0 & -1 & 0 & 1\end{pmatrix}$ & $v_2^\dagger = \begin{pmatrix} -1& 0& 1& 0\end{pmatrix}$  \\
   \bottomrule
  \end{tabular}
  \caption{The vectors needed in the {\em shrink} and {\em expand} stage for each direction. Since the vectors
  are column vectors, to save space, we show here the hermitian conjugates which are row vectors.}
 \label{tbl1}
\end{table}

The most complex part of the computation is the parallel transport
of the neighbors since it involves two matrix multiplications for
each direction. The time-consuming part is the color matrix
multiplication; the spinor matrix multiplication can be implemented
efficiently due to the special form of the spinor matrices. In fact,
the spinor multiplication is not computed directly; a standard trick
is used that effectively halves the amount of computation needed.
The key observation is that the spinorial matrices $(1\pm\gamma_\mu)$
are projection matrices on a two-dimensional space, i.e.
$(1\pm\gamma_\mu)= v_1 v_1^\dagger + v_2 v_2^\dagger$, where the two 
vectors depend on direction (see Table~\ref{tbl1}). We can then reduce 
the number of color-matrix color-vector multiplications from four to two. The 
calculation proceeds in three steps:
\begin{itemize}
\item{{\em Shrinking}: In Table \ref{tbl1}, for each of the eight
direction a pair of 4-vectors is given. We first compute two color
vectors: $\psi^{c}_{1,2} = \psi_n v_{1,2}$.} 
\item{{\em Color
multiplication}: Each of this vectors are then multiplied with $U$
producing two color vectors $U\psi^{c}_{1,2}$.} 
\item{{\em Expansion}:
the two vectors are combined together to produce the final result:
\beq
\psi_n' = U\psi^{c}_{1} v_1^{\dagger} + U\psi^{c}_{2} v_2^{\dagger}. 
\eeq
}
\end{itemize}
It is also worth pointing out that our choice for the $\gamma$-matrices 
makes the shrinking and expansion steps particularly simple:
there are no multiplications, only permutations and additions.

To compute $M(m;U)\bm{\psi}$ we need to repeat the steps above for each
lattice site.  At each site we need to load 9 spinor-vectors,
$\psi_n$ and $\psi_{n\pm\mu}$, and 8 color matrices, $U_\mu(n\pm\mu)$.
The final result, $(M(m;U)\bm{\psi})_n$, has to be stored back in the
memory. The total amount of data transported to and from memory is
$384$ floating point numbers representing $1536/3072$ bytes if
stored in single/double precision. The calculation at each site
requires $1368$ floating point operations. The ratio of bytes
transported per floating point operation is 1.12 in single precision
and 2.25 in double precision.

\section{GPU architecture and programming model}
\label{sec:3}

The methods presented in this paper were tested on NVIDIA GPUs and
implemented using the CUDA toolkit~\cite{CUDA}. The two architectures
that we used for our testing were the older GT200 and the newer
Fermi/GF100 architecture. In this section we will describe the
relevant features of the GPU architecture and the programming model.

\begin{figure}
\centering
\includegraphics[width=5.5in]{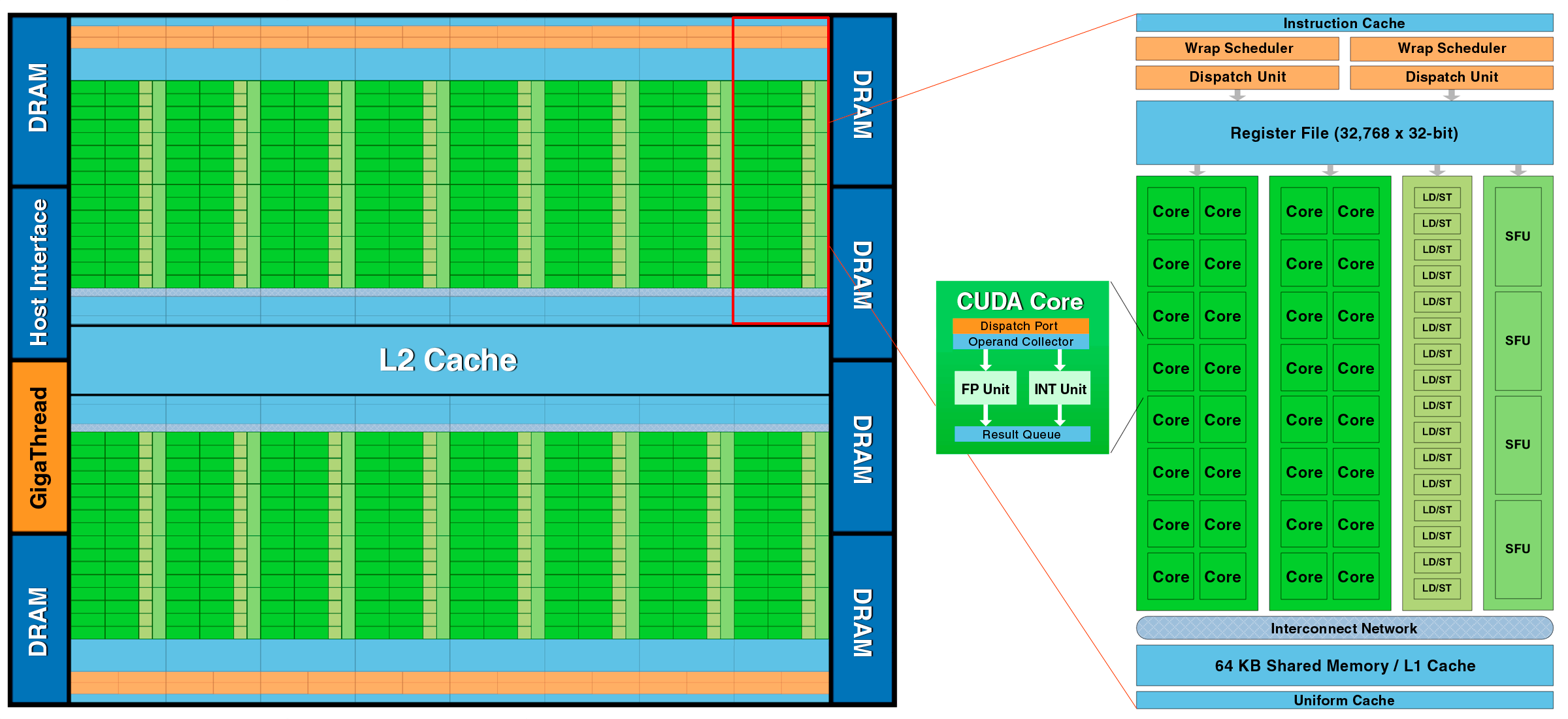}
\caption{Fermi architecture: 16 streaming multiprocessors (SM) 
each including 32 CUDA cores, 16 load/store units, 4 special 
functions units, a large register file, etc.~\cite{fermi-wp}}
\label{fig:fermi-arch}
\end{figure}

NVIDIA GPUs are designed to support a large number of threads running
in parallel. Typically, a GPU has a few hundred computing cores
each with one integer and one floating point unit. In single precision,
the floating point unit can execute one instruction per clock cycle.
Depending on the hardware architecture, the double precision performance
is one tenth to one half of the single precision performance. The routines executed
on the GPU compute cores are called {\em kernels} and they are issued
from the CPU. The CUDA toolkit allows for easy mixing of CPU code and 
kernels: the GPU code is identified, compiled separately and
linked in the final executable automatically.

The GPU architecture has a rich memory model: global device memory, 
L2 cache (in the case of Fermi), shared memory, etc. The majority of the device's storage
is global device memory. This is high-latency off-chip memory with a capacity
generally on the order of gigabytes. The CPU memory is accessible from the
GPU via the PCI bus which is more than 20 times slower
than the GPU's memory bus. In order to efficiently
use the GPU, frequently used data has to reside in device memory and
transfers to and from CPU memory have to be infrequent. In our codes
we store all the required data for the inversion in the device memory and
only transfer the result to the CPU memory when the inverter converges.

At the hardware level, the GPU has around a dozen streaming multiprocessors (SM),
each with 8-32 compute cores, a large register file and some shared memory.
The threads scheduled to run on one SM share the register file and shared 
memory. To run memory intensive applications efficiently, the GPU has a large
bandwidth to the device memory, around 150 GB/s. However, the device
memory latency is a few hundred clock cycles. To hide this latency the
SM keeps a pool of active threads that are scheduled to run as soon as
the required data is loaded from device memory. In order to hide the
latency completely, we need a couple of hundred threads to be active for
each SM. To quickly switch the active threads in and out of the execution 
queue, each thread gets exclusive access to its registers as long as the
thread is active. Thus, the register file needs to hold the registers for
hundreds of threads. This is why the number of registers per thread is limited
to 128 for the GT200 architecture and 63 for the Fermi architecture (the
number of registers per thread is reduced for Fermi since each SM has
four times as many cores whereas the number of registers was only doubled.)
When writing kernels we have to be very careful not to exceed this limit, since
any additional data will be spilled into device memory resulting in a significant
performance loss.

Threads are submitted to the processor divided into groups known as blocks.
A basic kernel call
\begin{verbatim}
kernel <<< block_number, block_size >>> (arguments ...);
\end{verbatim}
will launch $\texttt{block\_number} \times \texttt{block\_size}$ threads in blocks 
of size {\tt block\_size}. Simplifying things a little, we think of the kernel call above 
as launching many threads, each calling an identical function
\begin{verbatim}
kernel(arguments ..., thread_number);
\end{verbatim}
where the only difference is the {\tt thread\_number} provided by CUDA runtime.
Each block is mapped to a SM which enables cooperation of threads within their block. 
Threads within a block can quickly share data through their SM's shared 
memory and synchronize with block-wide barrier instructions. The shared
memory can also be used to store temporary data; this is very useful when
the number of registers needed for each thread exceeds the hardware limit
mentioned above. However, since shared memory size is actually smaller
than the register file (for Fermi we have 64KB of L1/shared memory vs
128KB for the register file), the additional amount of storage is limited. 
Furthermore, the data has to be carefully laid out in shared memory to
avoid bank conflicts (see section 5.3.2.3 of~\cite{CUDA}).

\begin{figure}
\centering
\includegraphics[width=5.5in]{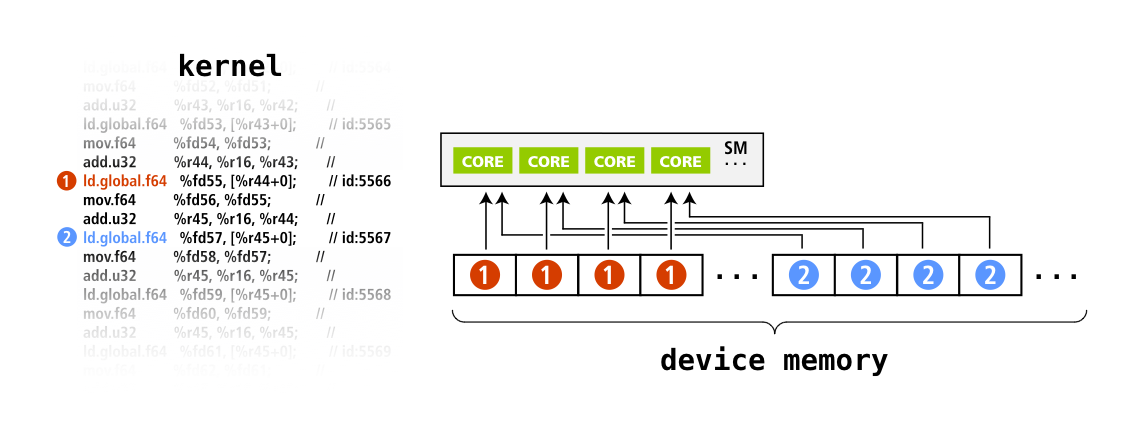}
\caption{Computing model: the {\em kernel} is a set of instructions executed in
parallel on all computing cores. When a load/store instruction is encountered, all
cores request a chunk of data from memory (usually register size). For best performance, 
these chunks should form a contiguous memory block.}
\label{fig:comp}
\end{figure}

Another important architectural detail is that, in order to fully exploit the
device memory bandwidth, the memory access needs to be {\em coalesced}.
This roughly means that threads within a {\em warp}, the smallest thread block that
gets scheduled for execution (32 for current devices), should access
data from a continuous block of memory. To make things clearer it is
useful to discuss the programing model we used to design our codes.
While the GPU programming model is more flexible, we find it useful to
model the GPU as a SIMD~\footnote{SIMD stands for single instruction,
multiple data.} device: each computing core executes the same set
of instructions on a different piece of data. We assume that each thread operates
independently, all threads execute the same set of instructions and each thread
loads and stores the input and output based on its thread index. To ensure
that the memory access is coalesced, we arrange the data loaded by the
same kernel instruction in a contiguous array (see Fig.~\ref{fig:comp}).
Thus the natural layout for our data is a {\em structure of arrays} rather than an
{\em array of structures}. For example an array of complex numbers will be
stored as an array with all real parts first and another array with the imaginary parts.

To sum up, in order to extract maximum performance from our kernels, we
need to store all relevant data in device memory, arrange it such that
the GPU cores that operate in lock-step load/store data in
continuos blocks and design kernels that use as few register as possible so that
we can schedule a large number of active threads to hide the latency of the 
device memory.

\section{Implementation details}
\label{sec:4}

In this section we discuss our GPU implementation of the Wilson matrix multiplication
routine. We discuss our implementation strategy, the data layout used for our vectors and gauge links 
and present the
multiplication kernel in detail paying close attention to the register usage.
We conclude this section with a discussion of the performance of the single
and double precision kernels showing that they nearly saturate the bandwidth limits. 

For lattice QCD simulations, most of the computation time is spent in the
routine that implements vector-matrix multiplication for the fermionic matrix.
Consequently, our efforts focused on optimizing the implementation of
this routine. This routine is well suited for parallel programming since
the steps described in Eqs.~\ref{eq:3},~\ref{eq:4} have to performed independently
for every site in the lattice. Moreover, each of the eight parallel transporters
described in Eq.~\ref{eq:4} can be performed independently. 

The most important design decision was the degree of thread granularity. On 
one hand, it is preferable to divide the computation in as many pieces as possible.
For example, we can have each thread compute only one parallel
transporter for one site. In this case, the threads would be light-weight and 
we can keep them active in sufficient numbers to hide memory latency. On the 
other hand, when a thread is responsible for a larger fraction of the computation,
the data is used more efficiently. For example, 
if each thread computes the matrix-vector multiplication for sites that belong to a 
$2^4$ tile, then the amount of data that needs to be moved per site is reduced to 65\%
of the data moved when computing one site per thread. Unfortunately, 
the register requirement per thread increases roughly 16 times. A possible solution
would be to use shared memory to load in the $2^4$ tile but this solutions adds
considerable complexity to the code. We decided that the optimal solution is to
have each thread responsible for one lattice site.

The layout of our data structures had to be carefully chosen to achieve maximum performance.
The vectors have the following structure:
\beq
\begin{split}
\bm{\psi} &= [ \overbrace{\text{color}_0, \text{color}_1, \text{color}_2}^{\text{real part}},
 \overbrace{\text{color}_0, \text{color}_1, \text{color}_2}^{\text{imaginary part}} ] \\
\text{color}_i &= [\text{spin}_0, \text{spin}_1, \text{spin}_2, \text{spin}_3] \\ 
\text{spin}_i &= [ \text{site}_0, \text{site}_1, \dots, \text{site}_N ] 
\end{split}
\label{eq:7}
\eeq
where $N$ is the number of sites on the lattice. We have then $2\times 3\times 4 = 24$
arrays, each of length $N$. 
The gauge links are represented by four $3\times 3$ matrices for each site.
Their layout is similar to the one used for vectors:
\beq
\begin{split}
U &= [ \overbrace{\text{dir}_x, \text{dir}_y, \text{dir}_z, \text{dir}_t}^{\text{real part}},
\overbrace{\text{dir}_x, \text{dir}_y, \text{dir}_z, \text{dir}_t}^{\text{imaginary part}}] \\
\text{dir}_i &= [\text{row}_0, \text{row}_1, \text{row}_2] \\ 
\text{row}_i &= [\text{column}_0, \text{column}_1, \text{column}_2] \\ 
\text{column}_i &= [ \text{site}_0, \text{site}_1, \dots, \text{site}_N ] 
\end{split}
\label{eq:8}
\eeq
As we discussed in the previous section, the data is arranged as a {\em structure
of arrays}. The particular order of these lattice-size arrays has little effect on the
performance; we could switch the spin and color indices within the vector layout
without penalty. On the other hand, if we rearrange the data so that the threads access
it with a stride the performance can easily decrease by an order of magnitude. 

The way the lattice sites are mapped in these linear arrays is not specified and, 
as long as the choice is made consistently, it can be tailored to our needs. 
For example, to implement even-odd preconditioning we 
choose a map that stores first the even parity sites and then the odd parity ones.
The matrix-vector multiplication routine uses this map to compute the position of the neighbors
in this linear array. To keep the routine flexible we can either use
layout routines to compute these indices on the fly, or precompute the
offsets of the neighbors and store them in eight arrays. We found that precomputing
the neighbors' offsets is the best strategy; it only adds $8\times 4=32$ bytes
to the data that needs to be transported per site which is a small overhead.

\begin{algorithm}[!t] 
\begin{algorithmic}[0]
\Procedure{dslash}{$\phi$, $\psi$, $U$, neighbors, $m a$, $i$} \Comment{$i$ is {\tt thread\_number}}
\State $\texttt{dest}[24]\gets 0,\:\texttt{spinor}[24]\gets 0,\:\texttt{half}[12]\gets 0$
\For{$dir$ = $\pm x,\dots,\pm t$}
\State $\texttt{spinor} \gets \psi[\text{neighbors}[dir][i]]$
\State $\texttt{half} \gets \textsc{shrink}(dir, \texttt{spinor})$
\For{$row$ = $0\:\textbf{to}\:2$}
\State $\texttt{spinor}_{0-5} \gets U(dir, i, row)$ 
\State $\texttt{spinor}_{12-23} \gets \textsc{color-mult}(\texttt{spinor}_{0-5}, \texttt{half}, row)$
\EndFor
\State $\texttt{dest} \gets \texttt{dest} + \textsc{expand}(dir, \texttt{spinor}_{12-23})$
\EndFor
\State{$\phi[i] \gets (ma + 4)\psi[i] - 0.5 \times \texttt{dest}$}\Comment{Use {\tt spinor} to load $\psi[i]$}
\EndProcedure 
\end{algorithmic} 
\caption{Matrix-vector multiplication: computes $\bm{\phi} = M(m; U) \bm{\psi}$}\label{dslash} 
\label{alg:1}
\end{algorithm}

A schematic presentation of the matrix-vector multiplication routine is presented
in Algorithm~\ref{alg:1}.  The subroutines $\textsc{shrink}$, $\textsc{color-mult}$ and
$\textsc{expand}$ are straightforward implementations of the steps described in
Section~\ref{sec:2}. It is important to note that, at a minimum, the kernel needs
space to store {\tt dest}, {\tt spinor} and {\tt half}, 60 numbers requiring 60/120
32-bit registers when using single/double precision. When using double precision
this poses a challenge. For the older GT200 architecture where the limit of 32-bit
register per thread is 128, the compiled kernel can almost fit completely in registers 
with very few temporaries stored in device memory. For the newer GF100 architecture
the register limit is 63 and there are too many temporaries stored in device memory -- 
the performance is reduced significantly. To work around this problem, we store
{\tt dest} in shared memory and {\tt spinor} and {\tt half} in the register file. 

\begin{figure}
\centering
\includegraphics[width=4.5in]{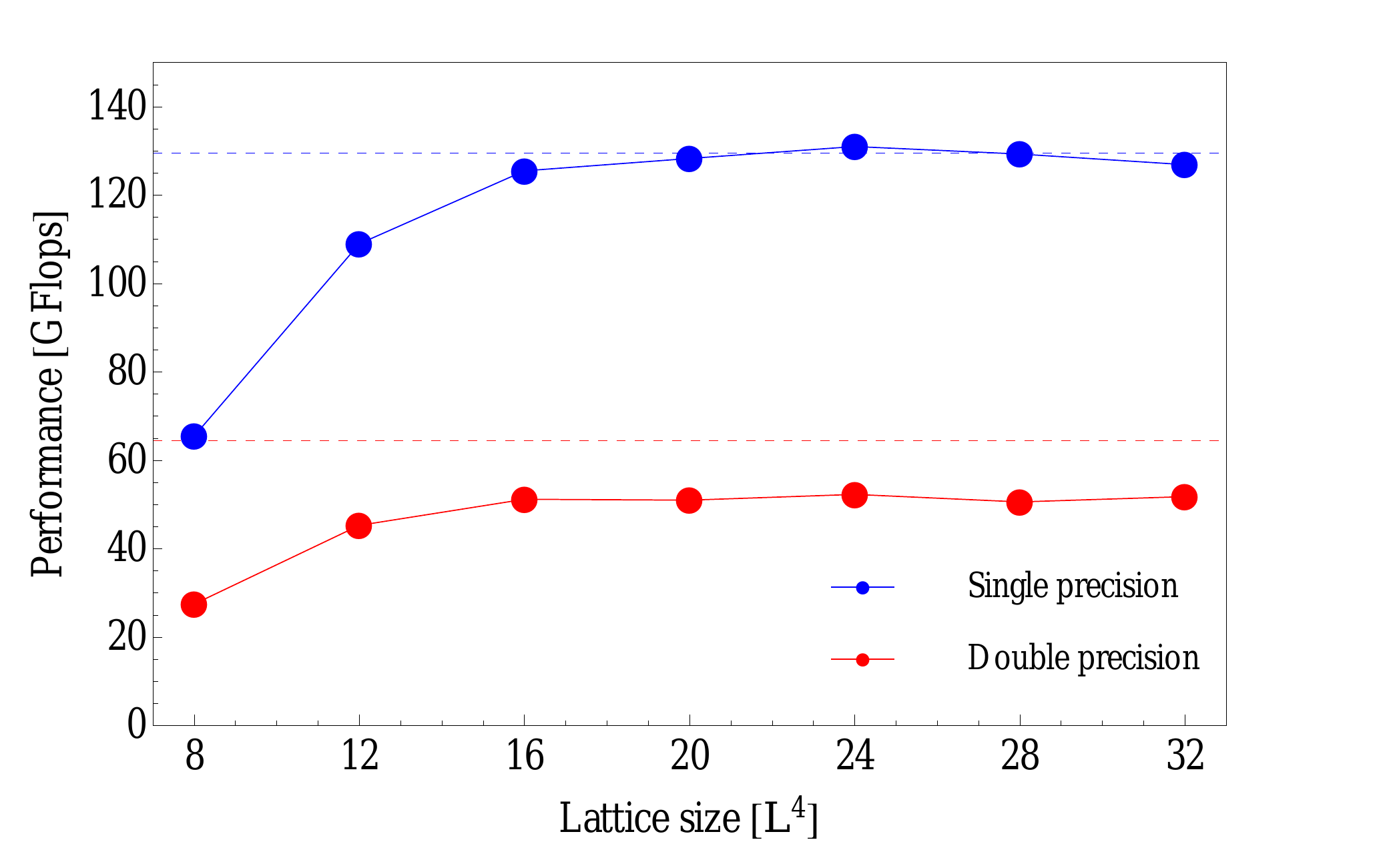}
\caption{{\sc dslash} routine performance on GTX 480 as a function of the lattice size. 
The dashed lines represent performance bounds imposed by bandwidth limits
(see text for details).}
\label{fig:dslash-perf}
\end{figure}

The performance of the {\sc dslash} routine as a function of the lattice size is 
presented in Fig.~\ref{fig:dslash-perf}. We see that for lattice sizes larger than 
$16^4$ performance saturates slightly above $50\,{\rm GFlops}$ for double 
precision kernels and around $130\,{\rm GFlops}$ for single precision. The
performance of our kernels is competitive with the performance of 
QUDA library~\cite{Clark:2009wm}, especially the double precision kernel. To gauge
the efficiency of our implementation, it is instructive to compare its performance
with the maximum allowed by the memory bandwidth. For GTX 480 the theoretical
bandwidth is $177.7\, {\rm GB/s}$, however this is not a very useful number since it is not
clear whether this bound can actually be achieved. A more useful bound is to take 
the peak bandwidth actually achieved by a very simple kernel, a vector addition
routine. This routine is perfectly parallel, its register requirement is minimal and the memory
access is completely sequential. As seen from Fig.~\ref{fig:vector-perf}, the maximal
bandwidth for this simple routine is about $145\, {\rm GB/s}$. In Section~\ref{sec:2}
we calculated the computational density for the matrix-vector multiplication routine to be
$1.12$ bytes/flop in single precision and $2.25$ in double precision. To reach a
bandwidth of $145\,{\rm GB/s}$ the performance of the kernels needs to be 
$129.5\,{\rm GFlops}$ in single precision and $65\,{\rm GFlops}$ in double precision.
It is easy to see from Fig.~\ref{fig:dslash-perf} that the single precision
kernel saturates this bound while the double precision kernel achieves 80\% of this
limit. The most likely reason is the fact that the maximum number of active threads that we
can schedule for this kernel is not sufficient to hide the memory bus latency.

We find that given the strategy we used, our matrix-vector multiplication kernels
are implemented nearly optimally. To improve their performance, we would need
to decrease the amount of data that needs to be moved from the device memory
to the GPU. This can be accomplished either by using a caching strategy or by 
employing a number of techniques to reduce the bandwidth: gauge fixing in the
temporal direction, using a compressed representation for the $SU(3)$ matrices,
using a different set of $\gamma$-matrices. The bandwidth-saving strategy was employed by
QUDA~\cite{Clark:2009wm}. For the single precision kernels, it was found that each 
of these techniques improves  the performance by about 10\% resulting in a significant 
improvement when used together. For double precision kernels the gains were more
modest. The only downside is that the code becomes considerably more complex.
We decided to keep our kernels simpler to make the transition to a multi-GPU framework 
smoother~\cite{gwu:2011}.

\section{Vector utilities}
\label{sec:5}

To implement our inverters, in addition to the matrix-vector multiplication routines we need
to implement utilities for vector operations: vector addition, multiplication with real and complex
scalars, scalar product and norm, etc. These routines are bandwidth limited since their computation
density is very low. The kernels are light-weight and simple to implement. To judge the implementation
efficiency it is more instructive to look at bandwidth utilization rather than floating point performance.

The problem with these kernels is that very often we need a combination of vector operations. For
example, when we want to evaluate $\bm{\psi}_3 \gets \bm{\psi}_1+\alpha \bm{\psi}_2$ we can either
break the computation in two steps, a vector multiplication and an addition, or write a new kernel. In the
first scenario the performance is suboptimal since a temporary vector, $\alpha\bm{\psi}_2$, is transported 
from the GPU to device memory and back. Since these kernels are bandwidth bound this 
severely impacts the code performance.
The other solution is to write a kernel for every linear operation, or at least for the ones used
most frequently. The problem with this approach is that it is error prone and very inflexible. When we
change the vector layout all these kernels have to be re-written.

To solve this problem we decided to use expression template techniques to generate the required
kernels on the fly~\cite{plagne:2007}. This technique uses the generic programing features offered by
templates in C++ to generate kernels at compile-time. The advantage is that we can write our algorithms
using clear mathematical syntax, the kernels are generated only when they are needed and the
code generated is optimal. The disadvantages are that the compilation time increases and the code
is brittle: a slight error causes many pages of errors that offer little information as to what caused the
compilation error. 

Our expression template implementation follows closely established 
techniques~\cite{plagne:2007}. Two details worth mentioning are the iterators used
and how we dealt with complex numbers. In order to implement expression templates 
for GPU vectors we needed a special class of iterators; these were provided by the 
Thrust library~\cite{Thrust}. As we pointed out in the beginning, most vector operations
traverse the arrays sequentially. However, multiplication with complex scalars is special:
the real and imaginary parts of the numbers at a given offset are needed. As discussed 
in the previous section, the layout of the GPU vectors stores the real and imaginary parts 
in separate arrays. To present a unified view of these arrays to our expression template
implementation, we always represent GPU arrays of complex numbers with the real part stored
in the first half of the array and the imaginary part in the second half; this is the reason for
having the real/imaginary index as the fastest varying index in the layouts described in 
Eq.~\ref{eq:7} and Eq.~\ref{eq:8}.

\begin{figure}
\centering
\includegraphics[width=4.5in]{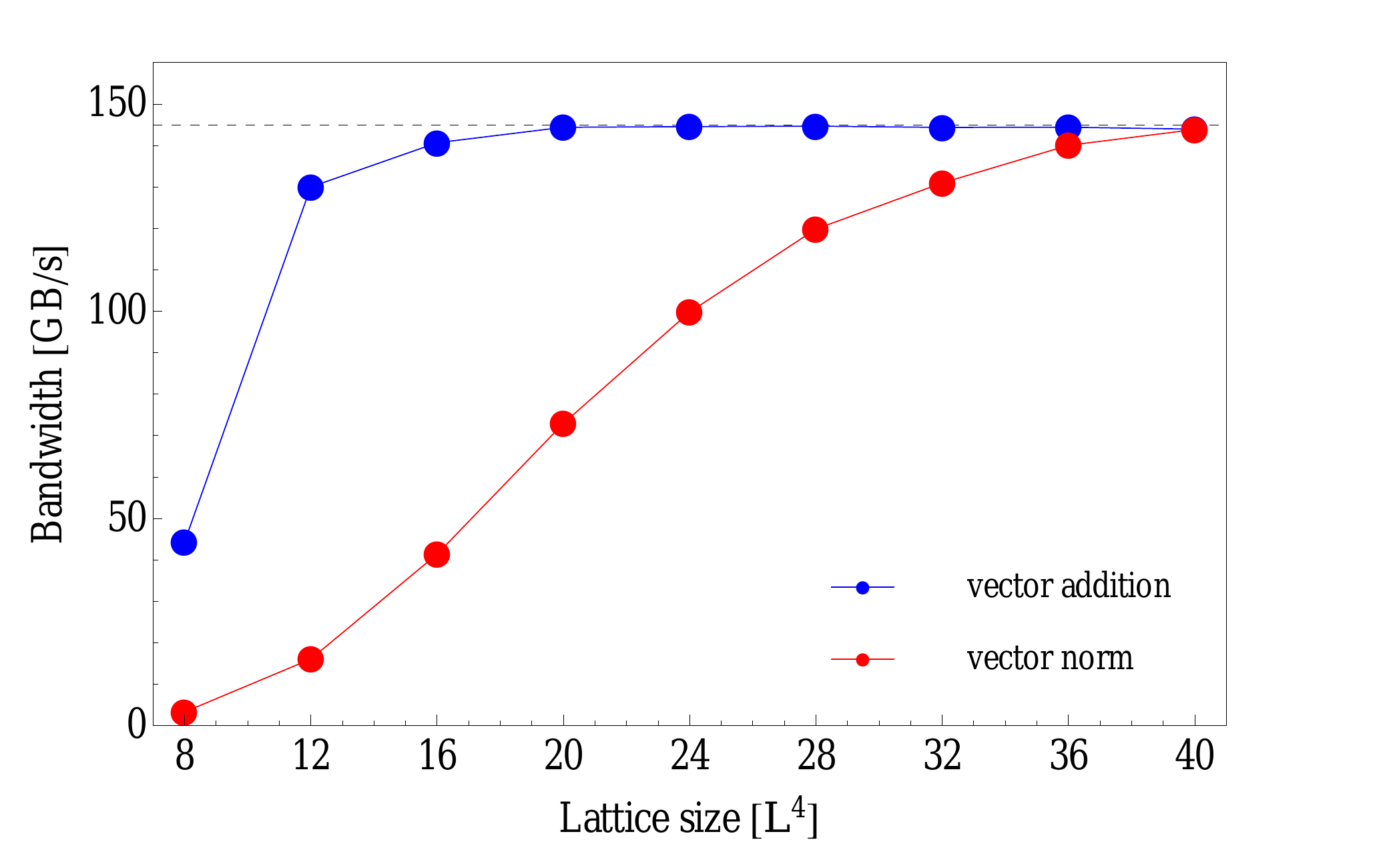}
\caption{Vector routine performance on GTX 480 as a function of the lattice size.
These routines are bandwidth bound and it is more informative to plot their bandwidth rather
than floating point performance.} 
\label{fig:vector-perf}
\end{figure}

In Fig.~\ref{fig:vector-perf} we present the performance of two kernels generated by the compiler,
the vector addition kernel and the norm calculation kernel. The vector addition kernel performance
is representative of all sequential access kernels; in particular, we measured the bandwidth for
$\bm{\psi}_3\gets \alpha\bm{\psi}_1+\beta\bm{\psi}_2$ with $\alpha$, $\beta$ being real, complex or zero.
The bandwidth is similar for all these operations. On the other hand, calculations of vector norm and 
scalar products are different. The data needs to be {\em reduced}, and this involves communication
which slows the kernel down affecting performance. We see that for lattices larger than $16^4$
the sequential kernels are saturating the device's memory bandwidth, whereas the reducing kernels need much
larger vectors to run optimally. It is important to notice that these very simple kernels peak at
about $145\,{\rm GB/s}$ bandwidth -- since they cannot be compute-bound and they are extremely
light-weight, we think that this offers a good measure of the GPU's maximal bandwidth.
Before we conclude, we want to mention that one of the steps in the BiCGstab algorithm, 
$\bm{\psi} \gets \bm{\psi} - \beta\bm{s} + \chi\bm{w}$, is different in structure from the kernels
tested above. We measured the performance of the automatically generated kernel for this step 
and we find it to be optimal (similar to the other sequential kernels), as expected.

\section{Multi-mass solvers}
\label{sec:6}

As mentioned in Section \ref{intro}, lattice calculations often require the solutions to linear 
systems
\be{
M(m_i; U) \bm\psi_i = \bm b\label{eq:6.1}
}
for several quark masses $m_i$. We will refer to such a set of linear equations as a 
\textit{multi-mass system} and any method to obtain the corresponding solutions as a 
\textit{multi-mass solver}. 

For practical reasons these systems are solved using iterative methods. A common choice is 
the so called Krylov inverters which iteratively build up the associated Krylov space, 
$\mathcal{K}_n(A,\bm{v})$, starting with some vector $\bm{v}$
\be{
\mathcal{K}_n(A,\bm{v})\equiv\text{span}\left\{ \bm v, A\bm v,A^2\bm v, \ldots, A^{n-1}\bm v\right\}
}
and seek the solution in this space by minimizing the residual~\cite{Saad:2003cy, Jegerlehner:1996pm}. 
These inverters can be used to determine the solution to Eq. \ref{eq:6.1} for a given quark mass. 
We refer to these methods as \textit{single-mass inverters}. Solutions to multi-mass systems can 
then be obtained by using the single-mass inverter for each mass separately.

There is another class of Krylov solvers designed to efficiently invert a set of shifted-linear systems,
i.e. a set of systems of the form
\be{
A(\sigma_i) \bm{x}_i = \bm{b} \quad \text{where} \quad A(\sigma_i) = A+\sigma_i\id \, ,
}
and $\sigma_i$'s are referred to as the shifts. The advantage of multi-shift inverters is that they use fewer 
matrix-vector multiplications compared to inverting each system individually. This is achieved by 
building only the Krylov space $\mathcal{K}_n\p{A,v}$ associated with the unshifted solution. 
The Krylov spaces for the shifted problems, $\mathcal{K}_n\p{A(\sigma_i),\bm{v}}$, 
are identical. For systems where the matrix-vector multiplication is significantly more
expensive than vector operations, these methods can be very effective.
From Eq.~\ref{eq:3}, it is clear that $M(m_i,U)$ can be cast 
in this form where each quark mass $m_i$ is associated with a shift $\sigma_i$. These methods are then 
well-suited for solving multi-mass problems. We will refer to methods of this type as 
\textit{multi-shift inverters}. 

One disadvantage of multi-shift inverters is that they do not lend themselves to certain optimizations
available to single-mass inverters. Moreover, they require for $n$ shifts the storage of $2n+1$ 
additional vectors \cite{Jegerlehner:1996pm}. Due to the current memory constraints on GPUs the extra 
storage required for the multi-shift inverters can be highly constraining. This constraint is also present 
in multi-node systems due to limited scalability. As a result, an optimized solver based on a single-mass 
inverter would be ideal, provided there is no substantial performance loss. This section is organized 
as follows: in Section \ref{sec:multi} we review some relevant details for multi-shift inverters. 
In Section \ref{single-shift} we discuss optimization techniques applicable to 
single-mass inverters: initial guess tuning, even-odd preconditioning and mixed-precision solvers.

\subsection{Multi-shift inverter}\label{sec:multi}

The convergence rate for an iterative inverter is determined by the conditioning number of the matrix. 
Systems with larger conditioning numbers will in general converge less quickly.
For multi-shift inverters it is important to choose the unshifted matrix to have the largest
conditioning number. Since the convergence is determined by the solution of the
unshifted system, a different choice would cause the inverter to exit prematurely.

For the fermionic matrix, the conditioning number is inversely proportional to the quark mass.
Therefore, systems with smaller quark masses will converge more slowly. We will abuse the notion of singularity slightly by referring to systems with smaller quark masses as being \textit{more singular}. In setting up the shifted systems we take the unshifted system to be the most singular system which we denote by $M(m_s; U)$ where $m_s=\min(m_i)$.\footnote{In the presence of interactions 
zero quark mass is actually achieved for $m=m_c < 0$. The most singular mass is the one closest to $m_c$.}
We then define the shifts $\sigma_i$ such that 
\be{
M(m_i; U) = M(m_s; U) +\sigma_i\id \qquad m_i\neq m_s
}
For $N$ quark masses we will have $N-1$ shifts. The multi-shift inverter then returns the solutions to
\be{
\br{M(m_s; U) +\sigma_i\id}\bm \psi_i = \bm b}
for all $\sigma_i$ as well as the unshifted solution $\bm{\psi}_0$. The inverter stops when 
the {\em residual} of the unshifted solution,
\be{
\bm r\equiv  \bm b - M(m_s; U){\bm\psi}_0 \, ,
}
satisfies the exit criterium 
$
\abs{\bm{r}}< \epsilon \abs{\bm{b}}
$ where $\epsilon$ is the desired accuracy. Since we set the unshifted matrix, $M(m_s; U)$, to the value of
the most singular mass, this exit criterion guarantees the algorithm, in exact arithmetic, will not exit 
before all solutions converge to the desired accuracy. Lastly, we note that less singular systems will 
converge more quickly, and it is therefore advantageous to stop updating them once they have converged. 
This can be achieved at virtually zero cost and is done in our implementation.

\subsection{Optimized single-mass inverter}\label{single-shift}

In this section, we discuss the implementation of an optimized single-mass inverter as a multi-mass solver. 
The single-mass based solver we implemented for this study utilizes even-odd preconditioning as well as mixed precision, neither of which can be utilized by multi-shift inverters. A third advantage of a single-mass inverter 
is the ability to make an initial guess for the solution. In this study we investigate two strategies for making initial guesses. One strategy relies on a polynomial extrapolation, and the other minimizes the residual 
in the space spanned by the solutions obtained for previous mass values. 


\subsubsection{Polynomial extrapolation}\label{poly}

In the polynomial extrapolation method, one uses the previous solutions to construct a polynomial from which the solution to the subsequent mass can be estimated by extrapolation. The estimated solution is then used as an initial guess. If the true function which maps $m$ values to their solutions is sufficiently smooth and does not vary rapidly
over the range being considered, this method is expected to generate a good guess. Specifically, 
suppose you have a set of values $m_1, m_2, \dots , m_N$ ordered such that $m_1 < m_2 < \dots < m_N$ and a set 
of previously determined solutions $\bm{\psi}_1, \bm{\psi}_2, ... , \bm{\psi}_k$.  From the set of solutions, one can
construct the polynomial 
\be{
\bm{\psi}(m)=\bm{c}_0+\bm{c}_1 m+\bm{c}_2 m^2+\dots+\bm{c}_k m^{k} \label{Poly}
}
with the coefficients $\bm{c}_0,\dots,\bm{c}_k$ chosen such that $\bm{\psi}(m_i)=\bm{\psi}_i$ for $i=1,\dots,k$.
We use this polynomial to make an initial guess $\bm{\psi}(m_{k+1})$ for the solution $\bm{\psi}_{k+1}$. To 
determine the coefficients, one must invert the associated Vandermonde's matrix
\be{
\left[\begin{array}{cccc}
1 & m_1 & m_1^2 &\\
1 & m_2 & m_2^2 &\dots \\
1 & m_3 & m_3^2&\\
&\vdots&&
\end{array}\right]
\left[\begin{array}{c}
\bm{c}_0\\
\bm{c}_1\\
\bm{c}_2\\
\vdots
\end{array}\right]
=
\left[\begin{array}{c}
\bm{\psi}_1\\
\bm{\psi}_2\\
\bm{\psi}_3\\
\vdots
\end{array}\right]
} 
The solution of such a linear system is well known, and is given by
\be{
\bm{\psi}(m)=\sum_j \bm{\psi}_j\prod_{i=1, i\neq j}^k\frac{m-m_i}{m_j-m_i}
}

\subsubsection{Solution space minimization}\label{min}

The second method considered is \textit{solution space minimization}. As suggested by the name, this method minimizes
the 2-norm of the residual in the space spanned by the solutions thus far obtained. This method is motivated by the
Krylov inverters which try to minimize the residual in the Krylov space. Instead of the Krylov space, we use the space
spanned by the solutions obtained for previous masses. This minimization process is desirable since it can be achieved
without any additional matrix vector multiplication. Its overall cost in practice is essentially zero.

To minimize the residual $\bm{r}$, one considers a general superposition of the previously determined solutions $\bm\psi_1, \bm\psi_2, ... , \bm\psi_m$ viz.

\be{
\bm{v}(\vec{c})=\sum_j c_j\bm{\psi}_j
}
When minimizing $\abs{\bm r}_2$ with respect to $\vec{c}$, one finds
\be{
\sum_jc_j\bm{d}_i^\dagger\bm d_j=\bm{d}^\dagger_i\bm b
}
where
\be{
\bm{d}_{k}=\bm b + (m-m_k)a \bm{\psi}_k,
}
and $m$ is the mass for which the solution is to be determined. The coefficients $c_j$ can then be determined readily
using standard numerical methods. 

\subsubsection{Even-odd preconditioning}\label{evenodd}

Even-odd preconditioning of the fermionic matrix speeds up the convergence of the inverter 
by a factor of 2 to 3. The even-odd preconditioning follows from first noticing that the fermionic 
matrix can be written as
\be{
M(m;U) = (ma+4)\mathds{1}-\frac{1}{2}
 \left(
 \begin{array}{cc} 
 0 & \dslash_{eo}(U)\\
  \dslash_{oe}(U) & 0 
  \end{array}
  \right)
}
when we separate the even and odd parity sites (the parity of a site $n$ is the
parity of $n_x+n_y+n_z+n_t$). In the following, we suppress the explicit gauge
field dependence. The linear system which needs to be solved can then be written as
\be{
\br{(ma+4)\mathds{1}-\frac{1}{2}
 \left(
 \begin{array}{cc} 
 0 & \dslash_{eo}\\
  \dslash_{oe} & 0 
  \end{array}
  \right)}
   \left(
 \begin{array}{c} 
 \bm{\psi}_{e}\\
  \bm{\psi}_{o} 
  \end{array}
  \right)
  =
    \left(
 \begin{array}{c} 
 \bm{b}_{e}\\
  \bm{b}_{o} 
  \end{array}
  \right)\label{eo}
}
which reduces to the set of equations 
\ba{
\br{(ma+4)^2\mathds{1}-\frac{1}{4} \dslash_{eo}\dslash_{oe}}\bm{\psi}_e&=&
(ma+4)\bm{b}_e+\frac{1}{2}\dslash_{eo}\bm{b}_o\label{eo2}\\
(ma+4)\bm{\psi}_o&=&\bm{b}_o+\frac{1}{2}\dslash_{oe}\bm{\psi}_e\label{eo3}
 }
The solution to Eq. \ref{eo} can then be found by first solving the preconditioned system 
$M_{\rm prec} \bm{\psi}_e = \bm{b}_{\rm prec}$ with
 \be{
 M_{\rm prec}=\br{(ma+4)^2\mathds{1}-\frac{1}{4} \dslash_{eo}\dslash_{oe}}\quad \text{and}\quad
 \bm{b}_{\rm prec}=(ma+4)\bm{b}_e+\frac{1}{2}\dslash_{eo}\bm{b}_o\label{eo4}
 }
for $\bm{\psi}_e$. The odd solution and hence the full solution can then be constructed using Eq. \ref{eo3}. 

\subsubsection{Mixed precision}\label{mix}

The efficiency of Krylov space inverters implemented on GPUs is limited by the memory 
bandwidth between the GPU main memory and processing units. As a result, single 
precision inverters will run roughly twice as fast as double precision inverters. However, 
lattice calculations often require higher accuracy than can be achieved in single precision.
A good compromise is offered by mixed-precision inverters~\cite{Clark:2009wm}. 
We used the {\em defect-correction} solver. To achieve the desired precision,
the bulk of the calculation is carried out in an inner loop by a fast, reduced-precision 
inverter to a lower tolerance. The residue and solution vectors are then updated in full precision,  
and the inner step is repeated until the desired precision is achieved. The results presented in this
paper were produced using double precision for the updating step and single precision in
the inner solver. All steps were run on the GPU.

\section{Results}\label{sec:res}

Our main goal in this work was to create a single-mass inverter that performs comparably to its 
multi-shift counterpart on GPUs. The underlying reason for this effort was the limited memory
of GPUs. As will be demonstrated in the following, for the cases considered in this study, an optimized 
single-mass inverter not only requires less memory than its multi-shift counter part, but also outperforms it. 
For the single-mass and multi-shift inverters we use BiCGstab and BiCGstab-M, respectively. 

\begin{table}[t]
\centering
\small\addtolength{\tabcolsep}{-3pt}
\begin{tabular}{lcccccccc}
\toprule
wide-range & -0.7908 & -0.7721 & -0.7511 & -0.7320 & -0.7105 & -0.6193 & -0.5108 & -0.2687 \\\midrule
narrow-range & -0.7908 & -0.7887 & -0.7866 & -0.7846 & -0.7825 & -0.7804 & -0.7784 & -0.7763 \\ 
	 \bottomrule
\end{tabular}
\caption{The two sets of masses considered in this study. The values listed here are $am$,
the masses in {\em lattice} units.
For the ensemble used in this study, the zero quark mass corresponds to $a m_c=-0.8173$. 
The {\em wide-ranged} set corresponds
to pion masses ranging from $500 \,{\rm MeV}$ to $2200 \,{\rm MeV}$ whereas the {\em narrow-ranged} set runs
from $500 \,{\rm MeV}$ to $600 \,{\rm MeV}$.}

\label{tab:kappas}
\end{table}

In order to test the
performance of the inverters, we use a small ensemble of 10 gauge configurations generated in the quenched
approximation using the standard Wilson action. The lattice spacing for these configurations is 
$a\approx 0.1\,{\rm fm}$. We use $24^4$ lattices which insures that the Wilson kernel and the 
sequential vector operations run at full speed (see Sections~\ref{sec:4} and \ref{sec:5}).
Considering the number of masses in our test sets, this is also the largest lattice size that can be used 
with multi-shift inverters on the GPUs available to us. For larger lattices the required vectors
will not fit in GPU memory. All results shown are averages taken with respect to this 
ensemble. The accuracy required for all solvers used in this section is $\epsilon=10^{-10}$.

We consider two sets of quark mass values: a wide-ranged set and a narrow-ranged set. The specific values 
can be found in Table \ref{tab:kappas}. The wide-ranged set is taken from a typical lattice QCD 
study~\cite{Alexandru:2009id}. To test the effectiveness of our solver we also used a narrow-ranged
set where the multi-shift inverter should have a large advantage over the single-mass one.

\begin{figure}[!t]
\centering
\includegraphics[width=5.5in]{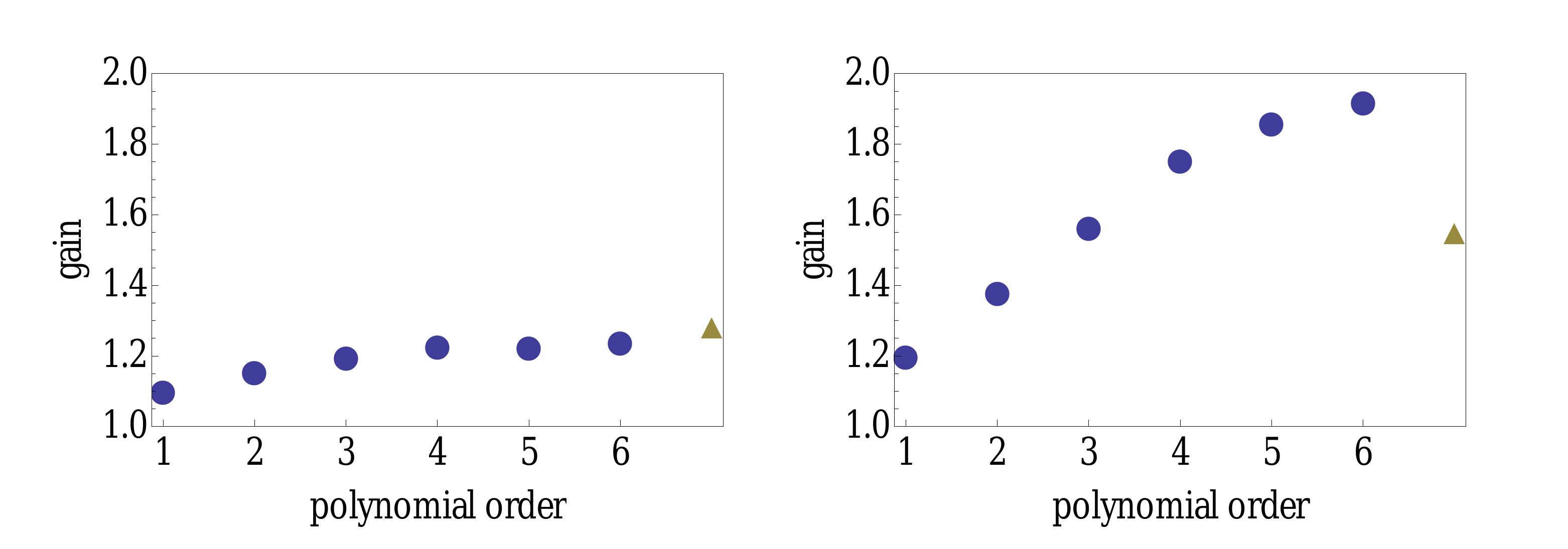}
\caption{Results for the guessing strategies for the two sets of mass values given in Table \ref{tab:kappas}:
left panel is the wide-ranged set and right panel is the narrow-ranged one. 
Blue circles and yellow triangles correspond to polynomial extrapolation and solution space minimization, 
respectively.  }
\label{results}
\end{figure}

We start by discussing the results for the initial guessing schemes.  
To quantify their effectiveness, we define the baseline to be the 
number of iterations required to solve the multi-mass system when taking the initial guess to be the null 
vector. We define the \textit{gain} to be
\be{
\text{gain}=\frac{\text{Total \# of iters.}}{\text{baseline}}
}
where the numerator represents the total number of iterations to solve the multi-mass system. 
As mention in Section \ref{single-shift}, the mass values are assumed to be ordered such that 
$m_1<m_2< \dots< m_N$, and the solutions are determined in order starting with $m_N$, the heaviest mass. 
We choose this order since we expect the guessing procedures to improve as the number of solutions increases. 
Since smaller masses require more iterations, better guesses for these masses will result 
in a larger reduction of the total number of iterations. 

We first consider the polynomial extrapolation. 
If the true function $\bm\psi(m)$ which maps mass values to their solution is slowly changing over 
the range of masses considered, then it should be well approximated by a finite order polynomial. 
It seems then sensible to use all available solutions and build the highest order polynomial
to extrapolate to the new mass. However, it is possible that using too high
an order might lead to oscillatory behavior (Runge's phenomenon). To investigate this, we limit the order
of the polynomial used in the extrapolation and measure the gain as we increase the polynomial order.
The results are shown in Fig.~\ref{results}. From this figure, we see that 
the largest gain is achieved when using the highest possible degree polynomial. We conclude that Runge's
phenomenon is not a concern and that we should use the highest polynomial order available.

The second guessing procedure considered is solution-space minimization. The results are presented 
in Fig.~\ref{results}. We see that the solution-space minimization is either comparable or worse than 
polynomial extrapolation. Moreover, polynomial extrapolation is simpler to implement. We therefore use 
the polynomial extrapolation method as our guessing strategy.

Our single-mass inverter uses a combination of even-odd preconditioning and mixed precision. As expected,
even-odd preconditioning reduces the number of iterations required for convergence by a factor of two.
Similarly, using a mixed-precision solver reduces the time per iteration by a factor of two.

\begin{table}[!t]
\centering
\begin{tabular}{llcccc} 
\toprule 
&solver& \text{Total iters} & \text{ms/iter} & \text{total time(s)} & \text{speed-up} \\
\midrule
\multirow{2}{*}{narrow-range} & \text{single-mass} & 1404 & 23.8 & 33.4 & \textbf{2.10} \\
&\text{multi-shift} & 790 & 88.9 & 70.2 & \\
 \midrule
 
\multirow{2}{*}{wide-range} & \text{single-mass} & 987 & 25.7 & 25.4 & \textbf{2.21} \\
&\text{multi-shift} & 793 & 70.8 & 56.1 & \\
 \bottomrule
\end{tabular}
 \caption{Performance results for the  two sets of masses. Here the speed-up is taken with 
 regard to the multi-shift inverter.}
  \label{tab:res}
 \end{table}

In the end, the important measure is the time-to-solution required by our solver. In Table~\ref{tab:res}, 
we compare the overall performance of the solver based on single-mass inverter to the one using 
a multi-shift inverter. We see that the single-mass inverter is faster than the multi-shift inverter 
by a factor of roughly two. While the number of iterations is larger for the single-shift inverter, the
total time per iteration is much smaller. On the one hand this is due to our using mixed-precision inverters.
On the other hand, the multi-shift inverter needs to carry out more algebra per iteration which slows it
down. This slowdown is reduced in the wide-range case because the shifted solutions converge quickly and
we stop updating them after few iterations.

On a related note,
it is worth pointing out that using an initial guess is more effective for the narrow-ranged set. This
is hardly surprising since in that case the masses are closer and the extrapolation works better.
The important fact is that using an initial guess helps the single-mass inverter exactly when this is
at a disadvantage with respect to the multi-shift inverter. In fact, if we didn't use an initial guess,
for the narrow-ranged set the single-mass inverter would be slower than the multi-shift one, whereas
for the wide-ranged set the single-mass inverter would still outperform it. Coupled with the slowdown
experienced by the multi-shift solver in the narrow-ranged case mentioned above, using an initial guess 
makes the single-mass inverter perform more uniformly across different scenarios.

We conclude that our single-mass solver is not just memory efficient 
but it actually outperforms the multi-shift inverter irrespective of the distribution of quark masses
considered.

\section{Conclusions}

In this study, we have addressed the issue of developing memory lean multi-mass solvers for use in 
lattice QCD calculations on GPUs. Such algorithms are crucial in order to carry out lattice calculations 
for the typical lattice sizes used today. We restricted our attention to the commonly used Krylov inverter 
BiCGstab and its multi-mass variant BiCGstab-M. However, we believe that our findings are applicable to 
other Krylov solvers such as CG and its multi-mass variant CG-M. 

We find that  a single-mass inverter using a combination of even-odd preconditioning, mixed-precision and a 
starting guess based on a polynomial extrapolation, is the best multi-mass solver. All three optimizations give comparable gains, each being responsible for approximately a factor of two speed-up. Thus, even though the 
multi-shift inverter requires significantly fewer iterations to converge, our single-mass inverter
actually outperforms it by more than a factor of two. Moreover, it uses significantly less memory.

It is worth pointing out that we only considered here mixed precision inverters based on single precision.
It was shown that half-precision inverters can be designed that increase performance by an additional 
50\%~\cite{Clark:2009wm}. This will make single-mass inverters outperform even more the multi-shift ones.

The motivation for our study was to develop inverters that perform well when constrained to fit in single
GPU memory. Lattice QCD codes are being developed that run on multiple GPUs and the memory constraint would
seem less relevant there. However, memory lean methods are also needed for multi-GPU codes due to scaling 
concerns. A memory lean solver requires fewer GPUs to accommodate a problem of the same size which
results in a more efficient use of resources.

We conclude with a remark about the relevance of our findings for codes running on CPUs.
Our conclusion should hold for all situations where the relative costs 
of different subroutines is the same. In the GPU case the timings are determined by the bandwidth requirements.
If Wilson fermion inverters running on CPUs are bandwidth limited too, then the single-mass inverter will
also perform better.

\section{Acknowledgements}

This work is partially supported by DOE grant DE-FG02-95ER-40907. We wish to thank Mike Clark, Ron Babich and
Balint Joo for useful discussions. The computational resources for this project were provided 
in part by the George Washington University IMPACT initiative. 

\bibliography{bicgstab_paper}
\bibliographystyle{model1-num-names}
  
\end{document}